\documentclass[conference]{IEEEtran}
\usepackage{csquotes}
\usepackage{cite}

\usepackage{graphicx}

\usepackage[cmex10]{amsmath}

\usepackage{algorithmic}

\usepackage{array}

\usepackage{mdwmath}
\usepackage{mdwtab}

\usepackage{eqparbox}
\usepackage{fixltx2e}

\usepackage{stfloats}
\usepackage{url}

\usepackage{caption}
\usepackage{multicol}
\usepackage{cite}
\usepackage{graphicx}
\usepackage{subfig}
\usepackage{bm}
\usepackage{CJK}
\usepackage{indentfirst}
\usepackage{amsmath}
\usepackage{eqnarray}
\usepackage{amssymb}
\usepackage{epstopdf} 
\usepackage{lipsum}
\newtheorem{theorem}{Theorem}[section]

\newcommand{\qed}{\nobreak \ifvmode \relax \else
	\ifdim\lastskip<1.5em \hskip-\lastskip
	\hskip1.5em plus0em minus0.5em \fi \nobreak
	\vrule height0.75em width0.5em depth0.25em\fi}


\makeatletter
\newcommand*\titleheader[1]{\gdef\@titleheader{#1}}
\AtBeginDocument{%
	\let\st@red@title\@title%
	\def\@title{%
		\bgroup\normalfont\large\centering\@titleheader\par\egroup
		\vskip0.5em\st@red@title}
}
\makeatother

\IEEEoverridecommandlockouts

\title{Performance Guaranteed Inertia Emulation for Diesel-Wind System Feed Microgrid via Model Reference Control}
\titleheader{2017 IEEE PES Innovative Smart Grid Technologies Conference, Washington D.C., USA}

\begin{document}
\author{
\IEEEauthorblockN{Yichen Zhang\IEEEauthorrefmark{1},
Alexander Melin\IEEEauthorrefmark{2},
Seddik Djouadi\IEEEauthorrefmark{1} and 
Mohammed Olama\IEEEauthorrefmark{3}}
\IEEEauthorblockA{\IEEEauthorrefmark{1}Department of Electrical Engineering and Computer Science, University of Tennessee, Knoxville, TN 37996-2250}
\IEEEauthorblockA{\IEEEauthorrefmark{2}Electrical \& Electronics Systems Research Division, Oak Ridge National Laboratory, Oak Ridge, TN 37831-6075}
\IEEEauthorblockA{\IEEEauthorrefmark{3}Computational Sciences and Engineering Division, Oak Ridge National Laboratory, Oak Ridge, TN 37831-6085}
\IEEEauthorblockA{Email: yzhan124@utk.edu, melina@ornl.gov, mdjouadi@utk.edu and olamahussemm@ornl.gov}
\thanks{Research sponsored by the Laboratory Directed Research and Development Program of Oak Ridge National Laboratory (ORNL), managed by UT-Battelle, LLC for the U.S. Department of Energy under Contract No. DE-AC05-00OR22725. The submitted manuscript has been authored by a contractor of the U.S. Government under Contract DE-AC05-00OR22725. Accordingly, the U.S. Government retains a nonexclusive, royalty-free license to publish or reproduce the published form of this contribution, or allow others to do so, for U.S. Government purposes.}}

\maketitle

\begin{abstract}
In this paper, a model reference control based inertia emulation strategy is proposed. Desired inertia can be precisely emulated through this control strategy so that guaranteed performance is ensured. A typical frequency response model with parametrical inertia is set to be the reference model. A measurement at a specific location delivers the information of disturbance acting on the diesel-wind system to the reference model. The objective is for the speed of the diesel-wind system to track the reference model. Since active power variation is dominantly governed by mechanical dynamics and modes, only mechanical dynamics and states, i.e., a swing-engine-governor system plus a reduced-order wind turbine generator, are involved in the feedback control design. The controller is implemented in a three-phase diesel-wind system feed microgrid. The results show exact synthetic inertia is emulated, leading to guaranteed performance and safety bounds.
\end{abstract}

%
\IEEEpeerreviewmaketitle

\section{Introduction}
Lack of inertia has been a crucial issue for microgrids under autonomous operation \cite{mgTrends} because most distributed energy resources (DER) are converter-interfaced and do not respond to frequency variations in the grid due to their decoupled control design. The solution is to implement supplementary loops to couple the  active power stored in converter-interfaced DER with rate-of-change of frequency (RoCoF), however, it is hard to assess how much synthetic inertia can be provided through this loop under disturbance, let alone emulate exact synthetic inertia. Under some specific control structures like droop control or virtual synchronous generator, the synthetic inertia can be estimated or controlled \cite{InertiaInDroop}, but this requires DER to operate as voltage sources and at the cost of de-loaded operation. On the another hand, guaranteed performance becomes necessary due to the increasing renewable penetration \cite{RampRates}. According to \cite{RampRates}, maintaining bounded frequency response under disturbance is a challenging control task.\par
Motivated by these issues, a novel inertia emulation strategy for converter-interfaced current-source DER is proposed. The model reference control (MRC) concept is employed \cite{gao2008network} to provide capability of emulating exact inertia. A diesel-wind system fed microgrid is used as a test system. A frequency response model, which is generally like a swing-prime-governor system, is defined as the reference model, where the desired inertia is parametrical. A measurement at a specific location delivers the information of disturbance acting on the diesel-wind system to the reference model. Then a static state feedback control law is designed to ensure the frequency of the physical plant tracks the reference model so that the desired inertia can be precisely emulated. Since active power variation is dominantly governed by mechanical dynamics and modes, only mechanical dynamics, i.e., the swing-engine-governor system plus a reduced-order wind turbine, are used in control design stage. Thus only mechanical states, which are easier to measure, are used in the feedback loop. By using this strategy exact synthetic inertia is emulated, leading to guaranteed performance and safety bounds.\par
The rest of the paper is organized as follows. Section \uppercase\expandafter{\romannumeral2} presents detailed model of the diesel-wind system. The reduced-order model of wind turbine generator is derived in Section \uppercase\expandafter{\romannumeral3}. The model reference control based inertia emulation strategy is expressed in Section \uppercase\expandafter{\romannumeral4}. Three-phase nonlinear simulation is illustrated in Section \uppercase\expandafter{\romannumeral5} followed by conclusion and future work in Section \uppercase\expandafter{\romannumeral6}.

\section{Diesel-Wind System Modelling}

\subsection{Diesel Generator}
The diesel generator model consists of diesel engine, speed governor, exciter, voltage regulator and a two-axis synchronous machine. The overall mathematical model is found in \cite{sauer1997power}. The diesel generator frequency response is governed by its mechanical dynamics and are expressed in Eq. (\ref{eq_frequency_response}): 
\begin{equation}
\label{eq_frequency_response}
\begin{split}
& M_{d}\Delta\dot{\omega}_{d} =\Delta P_{m}-\Delta P_{e,d}\\
& \tau_{d}\Delta\dot{P}_{m} = \Delta P_{v}-\Delta P_{m}\\
& \tau_{sm}\Delta\dot{P}_{v} =  -\Delta P_{v}  - (1/R_{d})\Delta\omega_{d}
\end{split}
\end{equation}
where $w_{d}$, $P_{m}$, $P_{v}$ are rotating speed, mechanical power and valve position, respectively. All parameters are scaled to microgrid level based on \cite{DieselData}.

\subsection{Type-4 Wind Turbine Generator}
In type-4 wind turbine generators (WTG), the permanent magnet synchronous generator (PMSG) is driven by the wind turbine and connected to a back-to-back voltage-source converter (VSC). The machine side converter (MSC) is used to regulate the speed of PMSG to achieve maximum power point tracking (MPPT), while the grid side converter (GSC) delivers the power in synchronous frequency. In this study, the averaged model of the converter is employed. The converter regulation is assumed to be infinitely fast so that:
\begin{equation}
\label{eq_cmd}
u_{q,\text{cmd}}= v_{q}, u_{d,\text{cmd}}= v_{d}
\end{equation}
where $v_{d}$ and $v_{q}$ are the converter output voltage in the d-q axis and the corresponding command values are denoted by the subscript \textquotedblleft cmd\textquotedblright.

\subsubsection{Wind Turbine}
The aerodynamic model can be represented as follows \cite{type4_yiwei}:
\begin{equation}
\label{eq_wind_power}
P_{t}=\frac{1}{2}\rho\pi R_{t}^{2}v_{\text{wind}}^{3}C_{p}(\lambda,\theta_{t})
\end{equation}
where
\begin{align}
	\label{eq_turbine}
	\lambda=& \frac{\omega_{t}R_{t}}{v_{\text{wind}}},\quad\lambda_{i}= \left(\frac{1}{\lambda-0.02\theta_{t}}-\frac{0.003}{\theta_{t}^{3}-1} \right)^{-1}\\
	C_{p}=& 0.73\left(\frac{151}{\lambda_{i}}-0.58\theta_{t}-0.002\theta_{t}^{2.14}-13.2\right) e^{-\frac{18.4}{\lambda_{i}}}
\end{align}
and $v_{\text{wind}}$, $R_{t}$, $\rho$, $\theta_{t}$ and $\omega_{t}$ are the wind speed, blade radius, air density, pitch angle of rotor blades and wind turbine speed, respectively. In this study, the WTG is assumed to operate at a partial loaded condition, so $\theta_{t}=0$ and pitch angle control is omitted. In addition, mechanical drive train is simplified using a constant gear ratio $k$. So the generator electric speed $\omega_{r}$, mechanical speed $\omega_{m}$ and the wind turbine speed $\omega_{t}$ have the following relationship $\omega_{r}=p\omega_{m}=pk\omega_{t}$, where $p$ is the pole pair number of the generator. The MPPT curve is given below \cite{type4_yiwei}:
\begin{equation}
\label{eq_mppt}
P_{\text{MPPT}}=C_{\text{opt}}\omega^{3}_{t}
\end{equation}
\subsubsection{Permanent Magnet Synchronous Generator and MSC Control}
The permanent magnet synchronous generator (PMSG) dynamics in the d-q axis as well as swing dynamics are given as follows \cite{type4_yiwei}:
\begin{equation}
\label{eq_pmsg}
\begin{split}
	&\dot{i}_{sd}=-\frac{R}{L_{d}}i_{sd}+p\omega_{m}\frac{L_{q}}{L_{d}}i_{sq}+\dfrac{1}{L_{d}}v_{sd} \\
	&\dot{i}_{sq}=-\frac{R}{L_{q}}i_{sq}-p\omega_{m}(\frac{L_{d}}{L_{q}}i_{sd}+\frac{1}{L_{q}}\varPsi)+\dfrac{1}{L_{q}}v_{sq} \\
	&\dot{\omega}_{m}=\frac{1}{M}(T_{e}-T_{m}-F\omega_{m})
\end{split}
\end{equation}
where ${i}_{sd}$, ${i}_{sq}$, ${v}_{sd}$, ${v}_{sq}$ are the stator current and voltage of PMSG. The electric torque is given as follows:
\begin{equation}
T_{e}=\frac{3}{2}p\left[\varPsi i_{sq}+(L_{d}-L_{q})i_{sd}i_{sq} \right] 
\end{equation}
The speed of PMSG is regulated by the MSC. The control diagram is shown in Fig. \ref{fig_Control_Torque}. The active power reference is given based on the MPPT curve calculated in Eq. (\ref{eq_mppt}). $u_{\text{ie}}$ is the supplementary control signal for inertia emulation. Defining the integrator outputs as three states $x_{1}$, $x_{2}$ and $x_{3}$, respectively, and substituting Eq. (\ref{eq_cmd}) yields the differential-algebraic model as follows:
\begin{figure}[t]
	\centering
	\includegraphics[scale=0.42]{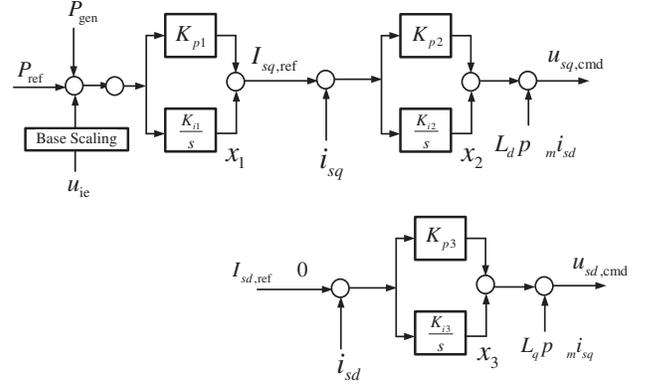}
	\caption{Rotor side converter control.}
	\label{fig_Control_Torque}
\end{figure}
\begin{align}
\label{eq_Control_Torque}
\begin{split}
\dot{x}_{1}=& K_{i1}(P_{\text{ref}}+u_{\text{ie}}-P_{e}) \\
\dot{x}_{2}=& K_{i2}\left[K_{i1}(P_{\text{ref}}+u_{\text{ie}}-P_{e})+x_{1}-i_{sq} \right] \\
\dot{x}_{3}=& -K_{i3}i_{sd}\\
0 =&-v_{sd}-L_{q}p\omega_{m}i_{sq}+x_{3}-K_{p3}i_{sd}\\
0 =&-v_{sq}+L_{d}p\omega_{m}i_{sd}+x_{2}\\
&+K_{p2}\left[K_{p1}(P_{\text{ref}}+u_{\text{ie}}-P_{e})+x_{1}-i_{sq} \right]\\
0=&-P_{e}+1.5(v_{sd}i_{sd}+v_{sq}i_{sq})
\end{split}
\end{align}

\subsubsection{Output L Filter Model and GSC Control}
Under the assumption of Eq. (\ref{eq_cmd}), the output filter model in the d-q axis is represented as follows 
\begin{equation}
\label{eq_filter}
\begin{split}
& \dot{i}_{ld}=-\frac{r_{f}}{L_{f}}i_{ld}+\omega i_{lq}-\frac{1}{L_{f}}v_{od}+\frac{1}{L_{f}}v_{gd}\\
& \dot{i}_{lq}=-\frac{r_{f}}{L_{f}}i_{lq}+\omega i_{ld}-\frac{1}{L_{f}}v_{oq}+\frac{1}{L_{f}}v_{gq}\\
\end{split}\end{equation}
where $v_{gd,q}$ is the GSC output voltage, ${i}_{ld,q}$ is current on the inductor and $v_{od,q}$ is the terminal voltage of converter connected to grid and assumed to be fixed. The standard P \& Q control in \cite{MG_control} is implemented, where the DC link voltage is regulated through active power control loop.

\section{Selective Modal Analysis Based WTG Model Reduction}
The selective modal analysis (SMA) based model reduction has been proved to be successful in capturing active power variation of WTG \cite{hector} and is chosen to achieve a reduced-order model. In WTG control design, the time frame of DC regulation is usually faster than MSC current loop for stability. Thus the DC link, GSC and output filter are simplified as a power flow-through and the corresponding dynamics are omitted in the model reduction.\par
Consider a type-4 WTG connected to a reference bus. Combining the equations from (\ref{eq_wind_power})--(\ref{eq_Control_Torque}) and linearizing them about the equilibrium points under $v_{\text{wind}}=12 (\text{m/s})$ yields the following state-space model:
\begin{subequations}
	\label{eq_linear_ss_full}
	\begin{align}
	\Delta\dot{x}_{W}&=A_{\text{sys}W}\Delta x_{W} + B_{\text{sys}W}\Delta u_{\text{ie}}\\
	\Delta P_{\text{gen}}&=C_{\text{sys}W}\Delta x_{W} +  D_{\text{sys}W}\Delta u_{\text{ie}}
	\end{align}
\end{subequations}
where the state vector is defined as $x_{W}=\left[i_{sd},i_{sq},\omega_{m},x_{1},x_{2},x_{3}\right]^{T}$ and the subscript $W$ denotes the WTG. The WTG rotor speed $\Delta\omega_{m}$ dynamics is closely related to its active power output, and the mode where $\Delta\omega_{r}$ has the highest participation would capture the most relevant active power dynamics. Therefore, $\Delta\omega_{m}$ is considered as the most relevant state, and the other states denoted as $z(t)$ are less relevant. Eq. (\ref{eq_linear_ss_full}) can be rearranged as
\begin{align}
\label{eq_linear_ss_arrange}
\begin{split}
\left[\begin{array}{c}\Delta\dot{\omega_{m}}\\\dot{z} \end{array}\right] &=\left[ \begin{array}{cc} 
A_{11} & A_{12}\\
A_{21} & A_{22}
\end{array} 
\right]\left[\begin{array}{c}\Delta\omega_{m}\\z \end{array} \right] 
+ \left[\begin{array}{c} B_{r}\\B_{z} \end{array} \right]u_{\text{ie}}\\
\Delta P_{\text{gen}}&=\left[C_{r} \quad C_{z} \right] \left[\begin{array}{c}\Delta\omega_{r}\\z \end{array}\right]
+D_{\text{sys}W}u_{\text{ie}}
\end{split}
\end{align}
The less relevant dynamics are:
\begin{equation}
\label{eq_linear_ss_less}
\dot{z}=A_{22}z+A_{21}\Delta\omega_{m} + B_{z}u_{\text{ie}}
\end{equation}
And the most relevant dynamic is described by: \\
\begin{equation}
\label{eq_linear_ss_more}
\Delta\dot{\omega}_{m}=A_{11}\Delta\omega_{m}+A_{12}z + B_{r}u_{\text{ie}}
\end{equation}
In (\ref{eq_linear_ss_more}), $z$ can be represented by the following expression:
\begin{equation}
\label{eq_sol_z}
\begin{split}
z(t)&=e^{A_{22}(t-t_{0})}z(t_{0})+\int_{t_{0}}^{t}e^{A_{22}(t-\tau)}A_{21}\Delta\omega_{m}(\tau)d\tau\\
&+\int_{t_{0}}^{t}e^{A_{22}(t-\tau)}B_{z}u_{\text{ie}} (\tau)d\tau
\end{split}
\end{equation}
Using the most relevant mode, $\Delta\omega_{m}(\tau)$ can be expressed as \cite{hector}: 
\begin{equation}
\label{eq_sol_r}
\Delta\omega_{m}(\tau)=c_{r}v_{r}e^{\lambda_{r}\tau}
\end{equation}
where $\lambda_{r}$ is the relevant eigenvalue, $v_{r}$ is the corresponding eigenvector and $c_{r}$ is an arbitrary constant. The accuracy of (\ref{eq_sol_r}) is guaranteed by the dominant term of $\Delta\omega_{m}$, which can be used in solving the first integral in (\ref{eq_sol_z}). Since $A_{22}$ is Hurwitz and its largest eigenvalue is much smaller than $\lambda_{r}$, the natural response will decay faster and can be omitted. The essential reason is that $A_{22}$ represents electrical dynamics which are faster than the electro-mechanical dynamic represented by $\lambda_{r}$. Then the response without control input in (\ref{eq_sol_z}) will be approximately equal to the forced response represented as follows:
\begin{align}
\label{eq_sol_integral_1}
e^{A_{22}(t-t_{0})}z(t_{0})+\int_{t_{0}}^{t}e^{A_{22}(t-\tau)}A_{21}\Delta\omega_{m}(\tau)d\tau\\
\approx (\lambda_{r}I-A_{22})^{-1}A_{21}\Delta\omega_{m}
\end{align}
The $u_{\text{ie}}$ is assumed to be fixed during the time window of interest, then the integral is calculated as
\begin{align}
\label{eq_sol_integral_rest}
&\int_{t_{0}}^{t}e^{A_{22}(t-\tau)}B_{z}u_{\text{ie}}(\tau)d\tau\approx (-A_{22})^{-1}B_{z}u_{\text{ie}}
\end{align}
Finally, the reduced-order WTG model with control inputs is
\begin{align}
\label{eq_linear_ss_reduced}
\begin{split}
\Delta\dot{\omega}_{m}&=A_{\text{rd}W}\Delta\omega_{m} + B_{\text{rd}W}u_{\text{ie}}\\
\Delta P_{\text{gen}}&=C_{\text{rd}W}\Delta\omega_{m} + D_{\text{rd}W}u_{\text{ie}}
\end{split}
\end{align}
where
\begin{align}
	\begin{split}
& A_{\text{rd}W}=A_{11}+A_{12}(\lambda_{r}I-A_{22})^{-1}A_{21} \\
& C_{\text{rd}W}=C_{r}+C_{z}(\lambda_{r}I-A_{22})^{-1}A_{21}\\
& B_{\text{rd}W}=B_{r} + A_{12}(-A_{22})^{-1}B_{z}\\
& D_{\text{rd}W}=D_{\text{sys}W} + C_{z}(-A_{22})^{-1}B_{z}
\end{split}
\end{align}

\section{Model Reference Control based Inertia Emulation Strategy}
\begin{figure*}[t]
	\centering
	\begin{minipage}[]{0.32\textwidth}
		\centering
		\includegraphics[scale=0.42]{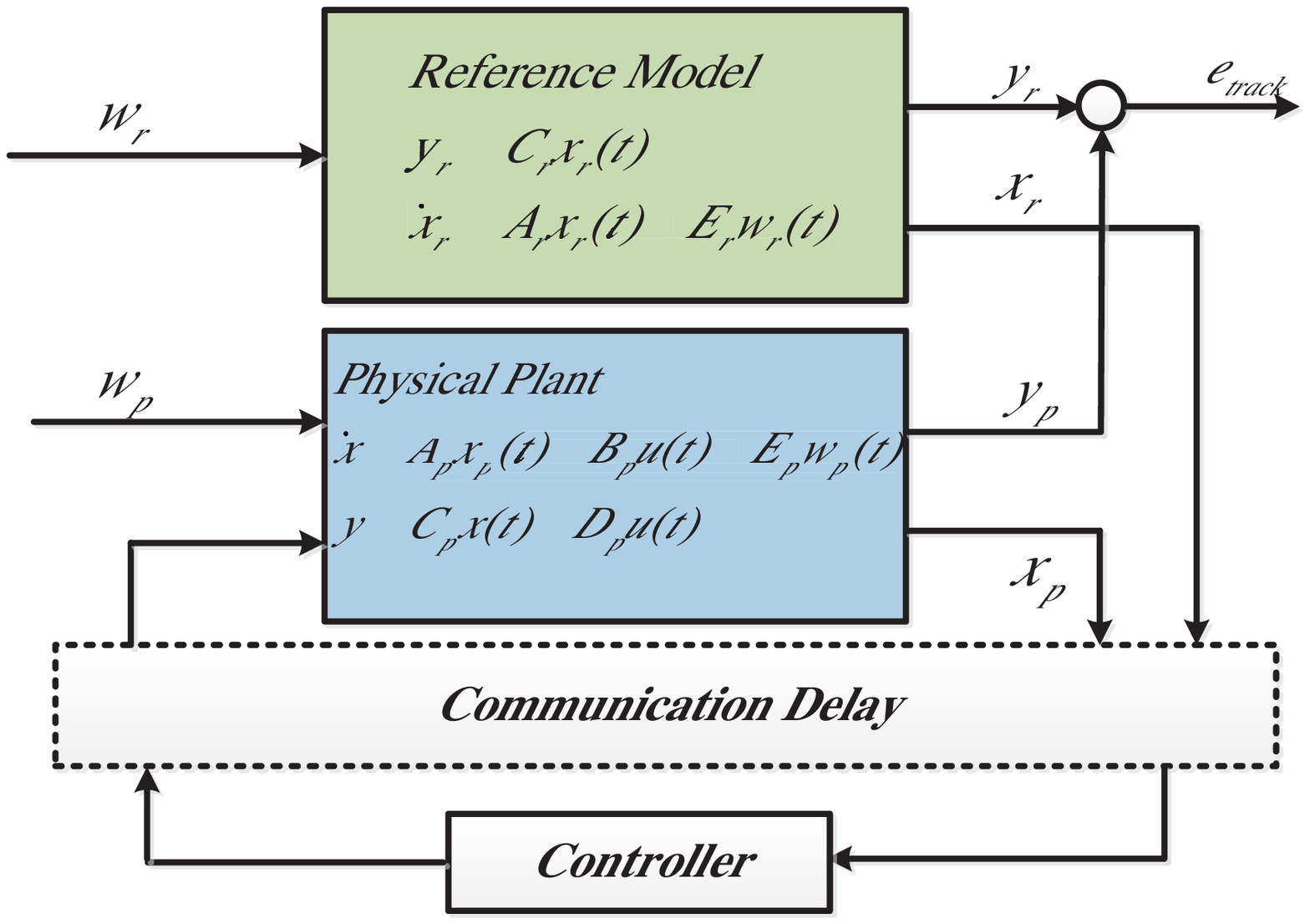}
		\caption{Typical model reference control structure.}\label{fig_mrc}
	\end{minipage}\hfill
	\begin{minipage}[]{0.55\textwidth}
		\centering
		\includegraphics[scale=0.38]{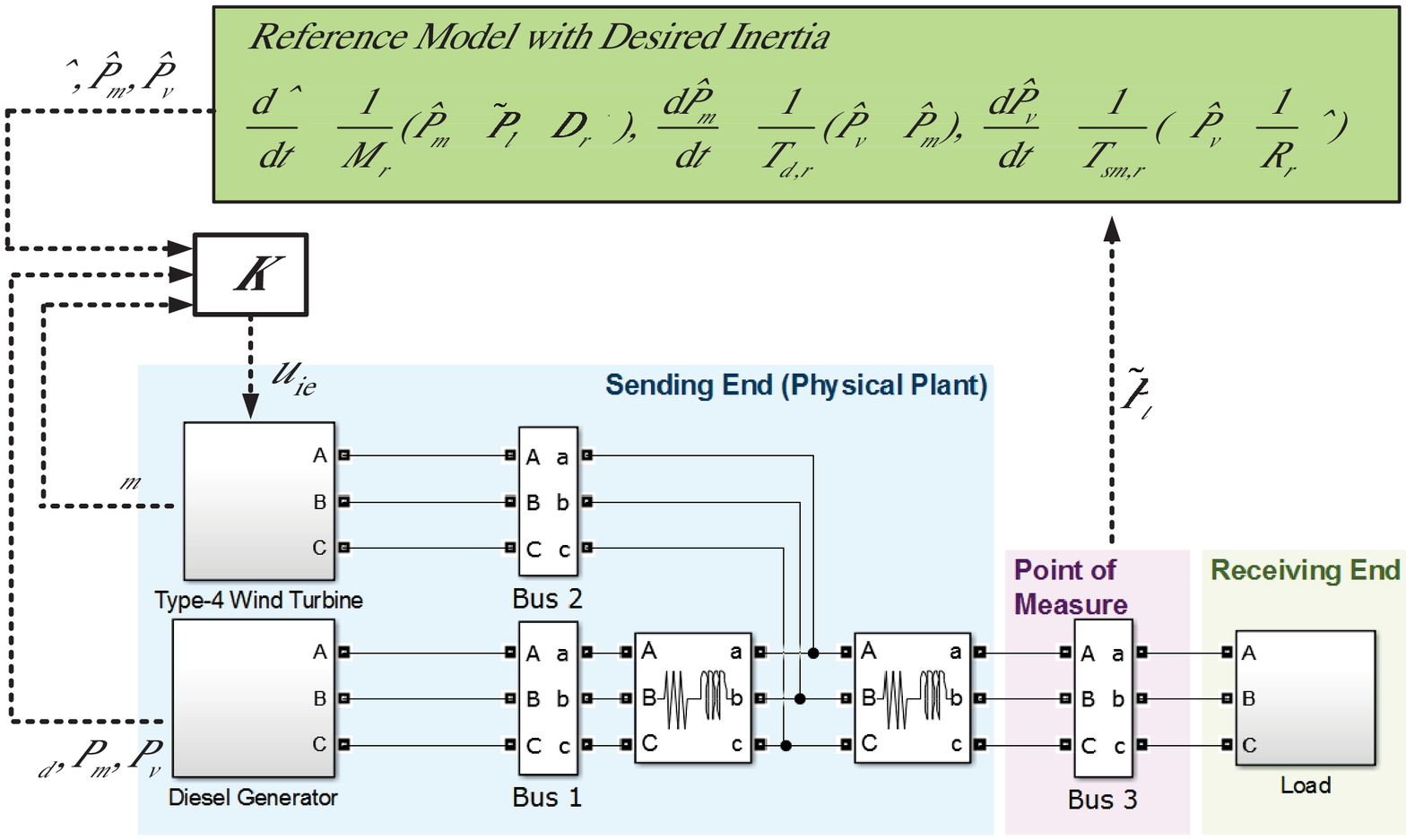}
		\caption{Diagram of model reference control based inertia emulation.}
		\label{fig_mrc_sim}
	\end{minipage}\hfill
\end{figure*}

The typical MRC structure is shown in Fig. \ref{fig_mrc}\cite{gao2008network}. The states of reference model $x_{r}$ and physical plant $x_{p}$ are measured. By closing the loop, the physical plant output $y_{p}$ will track the output of the reference model $y_{r}$. Fig. \ref{fig_mrc_sim} illustrates the MRC-based inertia emulation in our test system. The reference model is given as a frequency response model similar to Eq. (\ref{eq_frequency_response}) but with desired inertia $H_{r}$. The active power from the physical plant to the load is measured and the deviation value is sent to the reference model. It is worth mentioning that as shown in many previous studies \cite{hector}\cite{sfrm1990} the mechanical states and modes are enough to capture active power variations. So only mechanical dynamics are considered in the control design. It consists of the swing dynamics, diesel engine, speed governor expressed in Eq. (\ref{eq_frequency_response}) and the reduced-order model of wind turbine, which has been derived in the previous section. Thus only these states are measured in the feedback loop. As a result, the controller has been significantly simplified since these states are easier to measure and a state estimator is not necessary.  

\begin{figure*}[!b]
	\normalsize 
	\hrulefill
	\setcounter{equation}{21}
	\begin{equation}
		\label{eq_LMI_main}
		\left[ \begin{array}{ccccccccc} 
			\Theta_{11} & -\bar{U}_{1}+\bar{V}_{1}^{T} & \widetilde{B}\bar{K} & \bar{U}_{1} & 0 & \bar{E} & \bar{P}\bar{C}^{T} & \bar{P}\bar{A}^{T} & \bar{P}\bar{A}^{T} \\
			\ast & \Theta_{22}  & -\bar{U}_{2}+\bar{V}_{2}^{T} & \bar{V}_{1} & \bar{U}_{2} & 0 & 0 & 0 & 0                           \\
			\ast & \ast & -\bar{V}_{2}^{T}-\bar{V}_{2} & 0 & \bar{V}_{2} & 0 & \bar{K}D_{p}^{T} & \bar{K}\widetilde{B}^{T} & \bar{K}\widetilde{B}^{T} \\
			\ast & \ast & \ast & -\eta_{m}^{-1}\Upsilon_{1} & 0 & 0 & 0 & 0 & 0 \\
			\ast & \ast & \ast & \ast & -\kappa^{-1}\Upsilon_{2} & 0 & 0 & 0 & 0 \\
			\ast & \ast & \ast & \ast & \ast & -\gamma^{2}I & 0 & \bar{E}^{T} & \bar{E}^{T} \\
			\ast & \ast & \ast & \ast & \ast & \ast & -I & 0 & 0\\
			\ast & \ast & \ast & \ast & \ast & \ast & \ast & -\eta_{m}^{-1}\bar{M}_{1} & 0 \\
			\ast & \ast & \ast & \ast & \ast & \ast & \ast & \ast & -\kappa^{-1}\bar{M}_{2}
		\end{array} 
		\right]<0
	\end{equation}
	\begin{equation}
		\label{eq_Kov}
		K_{\text{mrc}}=\left[ \begin{array}{ccccccc}
			-15.22  & 3.90  &  3.89  &  9.21  & 13.85 & -7.90  & -3.37 \end{array}\right] 
	\end{equation}
	\setcounter{equation}{23}
	\vspace*{4pt}
\end{figure*}

Combining Eq. (\ref{eq_frequency_response}) and Eq. (\ref{eq_linear_ss_reduced}) yields the reduced-order model of the physical plant. The power flow is expressed as $\Delta P_{l}=\Delta P_{e,d}+\Delta P_{\text{gen}}$. The states, disturbance, input and output are defined as follows:
\begin{equation}
\label{eq_physical_define}
\begin{split}
& x_{p}=\left[\Delta\omega_{d},\Delta P_{m}, \Delta P_{v}, \Delta\omega_{m} \right]^{T}\\
& w_{p}=\Delta P_{l},u_{p}=u_{\text{ie}},y_{p}=\Delta\omega_{d}
\end{split}
\end{equation}
Similarly, the reference model is defined as:
\begin{equation}
\label{eq_reference_define}
\begin{split}
& x_{p}=\left[\hat{\omega}, \hat{P}_{m}, \hat{P}_{v} \right]^{T}\\
& w_{r}=\tilde{P}_{l},y_{r}=\hat{\omega}
\end{split}
\end{equation}
where $\Delta P_{l}=\tilde{P}_{l}$ and the corresponding equations are given in Fig. \ref{fig_mrc_sim}. Based on \cite{gao2008network} a state-feedback controller law is designed using linear matrix inequalities (LMI) which guarantee that the output of the closed-loop plant tracks the output of the reference model well in the $H_{\infty}$ sense. As illustrated both $x_{r}$ and $x_{p}$ are measured and the controller admits the following form
\begin{equation}
\label{eq_control_law}
u_{\text{ie}}=K_{p}x_{p}+K_{r}x_{r}
\end{equation}
Then the augmented closed-loop system is obtained as follow:
\begin{equation}
\label{eq_aug_ss}
\begin{split}
& \dot{x}_{ov}(t)=\bar{A}{x}_{ov}(t)+\bar{B}{x}_{ov}(t-\nu(t))+\bar{E}{w}_{ov}(t)\\
& e(t)=\bar{C}{x}_{ov}(t)+\bar{D}{x}_{ov}(t-\nu(t))
\end{split}
\end{equation}		
where		
\begin{align*}
& x_{ov}(t)=[x_{p}(t),x_{r}(t)]^{T},w_{ov}(t)=[w_{p}(t),w_{r}(t)]^{T}\\
& e(t)=y_{p}(t)-y_{r}(t),\bar{C}=[C_{p},-C{r}],\bar{D}=[D_{p}K_{p},D_{p}K_{r}]\\
& \bar{A}=\left[ \begin{array}{cc} A_{p} & 0\\ 0 & A_{r}\\ \end{array} \right],\bar{E}=\left[ \begin{array}{cc} E_{p} & 0\\ 0 & E_{r}\\ \end{array} \right]\\
& \bar{B}=\left[ \begin{array}{cc} B_{p}K_{p} & B_{p}K_{r}\\ 0 & 0\\ \end{array} \right]
\end{align*}		
Note that in Eq. \eqref{eq_aug_ss}, a time delay with the bound $ \eta_{m} \leq \nu(t) \leq \kappa$ is considered. Now we are in the position to introduce the controller design theorem.
\begin{theorem}
	\label{thm_main}
	Consider the system in \eqref{eq_aug_ss}. If there exists $\bar{P}>0$, $\bar{Q}>0$, $\bar{M}_{i}>0$ ,$\bar{U}_{i}$ ,$\bar{V}_{i}$ ,$i=1,2$, and $\bar{K}$ such that LMI in \eqref{eq_LMI_main} (at the bottom of the page) holds, where
	\begin{equation}
	\label{eq_LMI_1}
	\begin{split}
	&\widetilde{B}=[B_{p}^{T},0]^{T}\\
	&\Theta_{11}=\bar{A}\bar{P}+\bar{P}\bar{A}^{T}+\bar{Q}+\bar{U_{1}}^{T}+\bar{U_{1}}\\
	&\Theta_{22}=-\bar{Q}-\bar{V_{1}}^{T}-\bar{V_{1}}+\bar{U_{2}}^{T}+\bar{U_{2}}\\
	&\Upsilon_{i}=\bar{M}_{i}-2\bar{P},i=1,2
	\end{split}
	\end{equation}
   Then the state-feedback controller given in \eqref{eq_control_cal} can guarantee that the system in \eqref{eq_aug_ss} will attain output tracking performance $\gamma$ in $H_{\infty}$ sense:
   \begin{align}
	\label{eq_control_cal}
	[K_{p},K_{r}]=\bar{K}\bar{P}^{-1}
   \end{align} 
\end{theorem}\par

\section{Results and Discussion}
The wind turbine and PMSG parameters chosen can be found in \cite{type4_yiwei}. Other parameters are given as follows:
\begin{equation*}
\begin{split}
& M_{d}=4,M_{r}=6,R_{d}=R_{r}=0.03,D_{r}=0\\
&\tau_{d}=\tau_{d,r}=0.5,\tau_{sm}=\tau_{sm,r}=0.1
\end{split}
\end{equation*}
\begin{equation*}
\begin{split}
& K_{p1}=2^{-6},K_{i1}=10^{-6}\\
& K_{p2}=K_{p3}=0.8,K_{i2}=K_{i3}=0.5
\end{split}
\end{equation*}\par
\begin{figure}[h]
	\centering
	\includegraphics[scale=0.48]{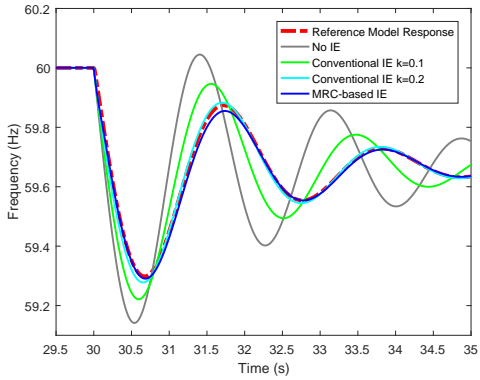}
	\caption{Diesel generator speed under MRC-based and conventional inertia emulation.}
	\label{fig_com_freq}
\end{figure}
\begin{figure}[h]
	\centering
	\includegraphics[scale=0.48]{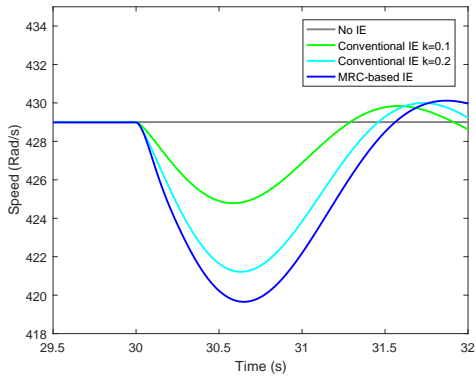}
	\caption{Wind turbine speed under MRC-based and conventional inertia emulation.}
	\label{fig_com_speed}
\end{figure}
\begin{figure}[h]
	\centering
	\includegraphics[scale=0.48]{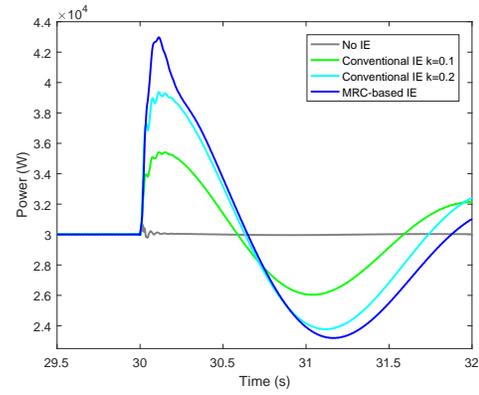}
	\caption{Wind turbine active power output under MRC-based and conventional inertia emulation.}
	\label{fig_com_power}
\end{figure}

The feedback control gain is presented in Eq. (\ref{eq_Kov}). The disturbance is considered as a step load change of 30 kW. The diesel generator speed is represented in Fig. \ref{fig_com_freq}. The actual speed (blue curve) tracks the virtual speed (red dash curve) generated by the reference model, where the desired inertia constant is set to be three seconds. By using the model reference control, one second synthetic inertia constant is precisely emulated. By having the guaranteed performance, safety bounds can be easily derived under the worst-case scenario. The PMSG speed and WTG active power variation are shown in Fig. \ref{fig_com_speed} and \ref{fig_com_power}, respectively.\par
The response under conventional inertia emulation realized by a washout filter $K_{w}s/(0.005s+1)$ with different gains is shown in Fig. \ref{fig_com_freq} as well. As seen, when $K_{w}=0.2$ the response is close to the one from reference model. However, a trial and error procedure is needed to reach desired performance, and will be sensitive to model uncertainty.

\section{Conclusion and Future Work}
In this paper, a novel model reference control based synthetic inertia emulation strategy is proposed. The reference model is designed to have a similar structure to the frequency response model with desired inertia. Through an active power measurement and state feedback, the wind turbine generator generates additional active power to guarantee that the diesel generator speed follow the frequency from the reference model. This novel control strategy ensures precise emulated inertia by the wind turbine generator as opposed to the trial and error procedure of conventional methods. By having guaranteed performance, safety bounds can be easily derived under the worst-case scenario. In addition, simultaneous emulation of inertia and damping coefficient can be realized. Moreover, inertia coordination of multiple renewable sources is capable via this control strategy as well.


\bibliography{IEEEabrv_zyc,Ref}  

\begin{thebibliography}{10}
\providecommand{\url}[1]{#1}
\csname url@samestyle\endcsname
\providecommand{\newblock}{\relax}
\providecommand{\bibinfo}[2]{#2}
\providecommand{\BIBentrySTDinterwordspacing}{\spaceskip=0pt\relax}
\providecommand{\BIBentryALTinterwordstretchfactor}{4}
\providecommand{\BIBentryALTinterwordspacing}{\spaceskip=\fontdimen2\font plus
\BIBentryALTinterwordstretchfactor\fontdimen3\font minus
  \fontdimen4\font\relax}
\providecommand{\BIBforeignlanguage}[2]{{%
\expandafter\ifx\csname l@#1\endcsname\relax
\typeout{** WARNING: IEEEtran.bst: No hyphenation pattern has been}%
\typeout{** loaded for the language `#1'. Using the pattern for}%
\typeout{** the default language instead.}%
\else
\language=\csname l@#1\endcsname
\fi
#2}}
\providecommand{\BIBdecl}{\relax}
\BIBdecl

\bibitem{mgTrends}
D.~E. Olivares, A.~Mehrizi-Sani, A.~H. Etemadi, C.~A. Ca{\~n}izares,
  R.~Iravani, M.~Kazerani, A.~H. Hajimiragha, O.~Gomis-Bellmunt, M.~Saeedifard,
  R.~Palma-Behnke \emph{et~al.}, ``Trends in microgrid control,'' \emph{{IEEE}
  Trans. Smart Grid}, vol.~5, no.~4, pp. 1905--1919, 2014.

\bibitem{InertiaInDroop}
S.~D'Arco and J.~A. Suul, ``Equivalence of virtual synchronous machines and
  frequency-droops for converter-based microgrids,'' \emph{{IEEE} Trans. Smart
  Grid}, vol.~5, no.~1, pp. 394--395, Jan 2014.

\bibitem{RampRates}
F.~M. Uriarte, C.~Smith, S.~VanBroekhoven, and R.~E. Hebner, ``Microgrid ramp
  rates and the inertial stability margin,'' \emph{{IEEE} Trans. Power Syst.},
  vol.~30, no.~6, pp. 3209--3216, Nov 2015.

\bibitem{gao2008network}
H.~Gao and T.~Chen, ``Network-based h-infinity output tracking control,''
  \emph{{IEEE} Trans. Autom. Control}, vol.~53, no.~3, pp. 655--667, 2008.

\bibitem{sauer1997power}
P.~W. Sauer and M.~Pai, \emph{Power system dynamics and stability}.\hskip 1em
  plus 0.5em minus 0.4em\relax New Jersey, USA: Prentice Hall, 1997.

\bibitem{DieselData}
G.~Kariniotakis and G.~Stavrakakis, ``A general simulation algorithm for the
  accurate assessment of isolated diesel-wind turbines systems interaction.
  part ii: Implementation of the algorithm and case-studies with induction
  generators,'' \emph{{IEEE} Trans. Energy Convers.}, vol.~10, no.~3, pp.
  584--590, 1995.

\bibitem{type4_yiwei}
Y.~Ma, L.~Yang, J.~Wang, F.~Wang, and L.~M. Tolbert, ``Emulating full-converter
  wind turbine by a single converter in a multiple converter based emulation
  system,'' in \emph{Proc. IEEE Applied Power Electronics Conference and
  Exposition (APEC)}, 2014, pp. 3042--3047.

\bibitem{MG_control}
J.~Rocabert, A.~Luna, F.~Blaabjerg, and P.~Rodriguez, ``Control of power
  converters in ac microgrids,'' \emph{{IEEE} Trans. Power Electron.}, vol.~27,
  no.~11, pp. 4734--4749, 2012.

\bibitem{hector}
H.~A. Pulgar-Painemal, ``Wind farm model for power system stability analysis,''
  Ph.D. dissertation, Univ. of Illinois at Urbana-Champaign, Champaign, IL,
  2010.

\bibitem{sfrm1990}
P.~M. Anderson and M.~Mirheydar, ``A low-order system frequency response
  model,'' \emph{{IEEE} Trans. Power Syst.}, vol.~5, no.~3, pp. 720--729, 1990.

\end{thebibliography}
\bibliographystyle{IEEEtran}

\end{document}